\documentclass[default]{sn-jnl}
\usepackage[utf8]{inputenc}

\theoremstyle{thmstyleone}%

\usepackage{amsmath}
\usepackage{graphicx}
\usepackage{lineno}

\newcommand{\beq}{\begin{equation}}
\newcommand{\eeq}{\end{equation}}
\newcommand{\beqa}{\begin{eqnarray}}
\newcommand{\eeqa}{\end{eqnarray}}
\newcommand{\beqann}{\begin{eqnarray*}}
\newcommand{\eeqann}{\end{eqnarray*}}
\bibpunct{(}{)}{;}{a}{}{,} 

\usepackage[normalem]{ulem}

\begin{document}

\title[\hspace{0cm}{Solar chromospheric heating and plasma outflows}]{Two-fluid numerical model of chromospheric heating and plasma outflows in a quiet-Sun}


\author*[1]{\fnm{K.} \sur{Murawski}}\email{kris.murawski@gmail.com}

\author[2,3]{\fnm{Z.E.} \sur{Musielak}}\email{zmusielak@uta.edu}

\author[4,1]{\fnm{S.} \sur{Poedts}}\email{stefaan.poedts@kuleuven.be}

\author[5]{\fnm{A.K.} \sur{Srivastava}}
\email{asrivastava.app@itbhu.ac.in}

\author[1]{\fnm{L.} \sur{Kadowaki}}\email{lkadowaki.astro@gmail.com}

\affil*[1]{\orgdiv{Institute of Physics}, \orgname{University of Maria Curie-Sk{\l}odowska}, \orgaddress{\street{Pl.\ Marii Curie-Sk{\l}odowskiej 5}, \city{20-031 Lublin}, 
\country{Poland}}}

\affil[2]{\orgdiv{Department of Physics}, \orgname{University of Texas at Arlington}, 
\city{Arlington}, 
\state{TX 7601}, \country{USA}}

\affil[3]{
\orgname{Leibniz-Institut fur Sonnenphysik (KIS)}, \orgaddress{\street{Schoeneckstr.\ 6}, \city{79104 Freiburg}, 
\country{Germany}}}

\affil[4]{
\orgdiv{Centre for mathematical Plasma-Astrophysics,
Department of Mathematics}, 
\orgname{KU Leuven}, \orgaddress{\street{Celestijnenlaan 200B}, \city{3001 Leuven}, 
\country{Belgium}}}

\affil[5]{
\orgdiv{Department of Physic}, 
\orgname{Indian Institute of Technology (BHU)}, 
\city{Varanasi-221005}, 
\country{India}}

\abstract{\textbf{Purpose:} This paper addresses 
long-standing solar physics problems, namely, 
the heating of 
the solar chromosphere and the origin of the solar wind. 
Our aim is to reveal the related mechanisms behind 
chromospheric heating and plasma outflows in a quiet-Sun. 
 \textbf{Methods:} 
The approach is based on a two-fluid numerical model 
that accounts for thermal non-equilibrium (ionization/{re\-com\-bi\-na\-tion}), 
non-adiabatic and non-ideal 
dynamics of protons+electrons 
and hydrogen atoms. 
The model is applied to numerically simulate the 
propagation and dissipation of granulation-generated waves 
in the chromosphere 
and plasma flows inside 
a quiet region. 
 \textbf{Results:} 
The obtained results 
demonstrate that collisions between protons+electrons and 
hydrogen atoms 
supplemented by plasma viscosity, 
magnetic resistivity, and recombination 
lead to thermal energy release, which 
compensates radiative and thermal losses in the chromosphere, 
and sustains the atmosphere with 
vertical profiles of averaged temperature 
and periods of generated waves that 
are consistent with recent observational data. 
 \textbf{Conclusion:} 
Our model conjectures a
most robust and global physical picture of granulation generated wave motions, plasma flows, and subsequent heating, which form and dynamically couple the various layers of the solar atmosphere. 
}

\keywords{Methods: numerical -- Sun: atmosphere -- Sun: activity}
%
%
%
\maketitle

%
%
\section{Introduction}
One of the major, long-standing problems of solar physics concerns 
the source of the thermal energy required to heat the different layers of the atmosphere. 
Observations show that the atmosphere, with the more than one million 
Kelvin hot solar corona, efficiently radiates its energy and thus it must be 
heated to be maintained in its quasi-stationary state. 
For decades, different heating mechanisms were proposed but so far no 
common agreement regarding a complete quantitative and qualitative picture has been reached.  In other words, the 
main physical process(es) underlying this heating of the different atmospheric layers still remains unknown. Space-borne and ground-based observational data stimulated various plausible explanations 
for this heating problem, among which a wave generation and dissipation mechanism is promising, especially for the chromosphere.  The latter 
mechanism is based on thermal energy deposition essentially by compressible waves. 
Such compressible waves are generated by turbulent motions occurring in the solar convective zone and by granular motions in the photosphere, since both regions are vast reservoirs 
of mechanical energy that can be converted into wave motions. 

The role of compressible 
(acoustic) waves in the solar chromospheric 
heating problem was first recognized by 
\cite{Biermann1946} and \cite{Sch1948}. Contemporary high-resolution observations 
revealed with unprecedented spatial and temporal resolution that the presence of 
different types of waves and flows constitutes an integral part of the solar atmosphere \citep[e.g.,] [] 
{Hansteen2010ApJ...718.1070H,Tian2011ApJ...738...18T,Dadashi2011A&A...534A..90D,DeMoortel2012,Srivastavaetal2017,Kayshap2015,Kayshap2020,Tian2021arXiv210202429T}. 
The excitation and propagation of magneto-acoustic waves 
driven by the granulation have been investigated by many authors 
\citep[e.g.,][]{Hansteen2006ApJ...647L..73H,Heggland2011ApJ...743..142H}. 
Additionally, 
\cite{Hansteen2010ApJ...718.1070H} 
and 
\cite{2022arXiv220702878F}
showed that 
the transition region and coronal heating results from the buffeting of 
the magnetic field lines by turbulent motions in the photosphere and 
in the convection zone as well as from 
the injection of emerging magnetic flux. 

The dissipation process is more difficult to address, but multiple works 
\citep[e.g.,][]{Martinez-Sykoraetal2017,Snow2021MNRAS.506.1334S,Wang2021ApJ...916L..10W}  studied the wave dissipation by the shock wave and/or non-ideal MHD effects, including ion-neutral collisions. 
Specifically, \citet{Martinez-Sykoraetal2017} studied the excitation of solar spicules by 
the solar granulation and their 2.5-dimensional (2.5D) model developed within the framework of 
radiative magnetohydrodynamics (MHD) considered a partially-ionized solar plasma and modelled 
its neutrals by ambipolar diffusion. In two other more recent studies, \citet{Fleck2021RSPTA.37900170F}
performed numerical simulations of acoustic-gravity waves that were excited by the solar 
granulation, and \citet{Snow2021MNRAS.506.1334S} used a two-fluid model to investigate the 
role of slow shocks in the solar atmosphere but without taking into account the solar 
granulation. In the approach presented in this paper, we develop a two-fluid 2.5D numerical 
model that accounts for interaction of protons with hydrogen atoms, and includes radiative
loss terms, whose effects of our simulations and the obtained results are studied and 
discussed. 

Nowadays, numerical simulations play a complementary role to observations in exploration 
of the solar atmosphere and in particular in understanding the propagation of waves and their 
contribution to the chromospheric and coronal heating. 
In this context, \citet{khomenko2012ApJ...747...87K} 
studied the heating of the solar chromosphere 
resulting from ion-neutral collisions, referred to as 
ambipolar diffusion. The authors concluded that ambipolar diffusion 
has the potential to efficiently heat the chromosphere. 
Additionally, 
\citet{Kuzmaetal2019a}, 
\citet{2021A&A...652A.124N}, 
and 
\citet{2021A&A...652A.114P} 
showed that in the regime of two-fluid, 
respectively monochromatic acoustic, impulsively generated 
magneto-acoustic and Alfv\'en 
waves 
are likely to effectively heat the chromosphere.  
Moreover, \citet{Srivastavaetal2018} proposed that 
the small-scale, two-fluid penumbral jets that are omnipresent in active regions, possess sufficient energy to heat the 
solar corona. 
In other papers, \citet{Wojciketal2018,Wojciketal2019c} 
confirmed that 
ion-neutral collisions result in thermal energy release 
\citep{Martinez-Sykora2020ApJ...889...95M,Carlssonetal2019ARA&A..57..189C}. 
\citet{Manevaetal2017} demonstrated 
that two-fluid ion magneto-acoustic-gravity waves locally heat solar magnetic flux-tubes.  
\citet{Wojciketal2019a} and \citet{2020ApJ...896L...1M} 
performed respectively 2D and 3D radiative numerical simulations of 
granulation generated two-fluid waves that effectively heat the plasma, 
compensating for the radiative energy losses. 
\citet{Wojciketal2019b} showed that granulation-generated jets 
and associated plasma 
outflows may contribute to the origin of the fast component of the solar wind \citep{Tianetal2014}. 
There are also recent studies that associate the network jets with propagating heating events and not strong flows  \citep[e.g.,][]{DePontieuetal2017}. 

Despite the above studies, which 
addressed some parts of the localized heating 
problem, a full treatment of the energy flow from the deeper and cooler to the outer and hot 
solar atmospheric layers still remains unsolved.  
Yet, such full treatment including all these layers is necessary in order
to solve this heating problem in 
the solar chromosphere.  
Therefore, this paper is devoted to such 
a general approach in which the problem is addressed by studying the propagation and 
dissipation of granulation generated waves and plasma flows in a self-consistent way.  
More specifically, 
the earlier two-fluid models of 
\citet{Wojciketal2019b,Wojciketal2019c,Wojciketal2019a} 
and \citet{2020ApJ...896L...1M} 
are 
generalized by considering all non-adiabatic and non-ideal effects as well as ionization 
and recombination within the two-fluid model of 
the solar atmosphere, 
which takes into account collisions between protons+electrons and neutrals (hydrogen atoms). 
The following section presents a detailed description of the model. 
Sections~3 and 4 contain 
the numerical results and the conclusions, respectively. 
\section{Physical model and governing equations}
\label{sec:num_model}
Consider 
the solar atmosphere as a system consisting of 
interacting fluids: 
ions (protons+electrons) and hydrogen atoms, 
denoted respectively by subscripts $_{\rm i}$ and $_{\rm n}$. 
Each fluid is characterized by its number density $n_{\rm k}$, 
${\rm k}=\{{\rm i}, {\rm n}\}$, 
mass density $\varrho_{\rm k}=n_{\rm k}m_{\rm k}$ with 
mass $m_{\rm k}$, velocity $\mathbf{V}_{\rm k}$, 
gas pressure $p_{\rm k}$, and temperature $T_{\rm k}$. 
These fluids are 
described by the following equations 
\citep[e.g.,][]{Zaqarashvilietal2011,khomenko2012ApJ...747...87K,Meier2012,2014SSRv..184..107L, 
Oliveretal2016,Manevaetal2017,PopescuBraileanuetal2019}: 
\begin{equation}
\frac{\partial \varrho_{\rm i}}
{\partial t}+\nabla\cdot(\varrho_{\rm i} \mathbf{V}_{\rm i}) = 
m_{\rm i} (\Gamma^{\rm ion}_{\rm i}+\Gamma^{\rm rec}_{\rm i}) 
\, , 
\label{eq:ion_continuity} 
\end{equation}
\begin{equation}
\frac{\partial \varrho_{\rm n}}{\partial t}+
\nabla\cdot(\varrho_{\rm n} \mathbf{V}_{\rm n}) = 
m_{\rm n} (\Gamma^{\rm ion}_{\rm n}+\Gamma^{\rm rec}_{\rm n}) 
\, , 
\label{eq:neutral_continuity} 
\end{equation}
\begin{multline}
\frac{\partial  (\varrho_{\rm i} \mathbf{V}_{\rm i})}{\partial t}
+
\nabla \cdot (\varrho_{\rm i} \mathbf{V}_{\rm i} \mathbf{V}_{\rm i}
+
p_{\rm i}\mathbf{I}) 
= 
\varrho_{\rm i} \mathbf{g} 
+ 
\frac{1}{\mu}(\nabla \times \mathbf{B}) \times \mathbf{B} 
+ 
\nabla\cdot\mathbf{\Pi}_{\rm i} 
+ 
\mathbf{S_{\rm i}}
\, ,
\label{eq:ion_momentum}
\end{multline}
\begin{multline}
\frac{\partial (\varrho_{\rm n} \mathbf{V}_{\rm n})}{\partial t} 
+
\nabla \cdot (\varrho_{\rm n} \mathbf{V}_{\rm n} \mathbf{V}_{\rm n}
+
p_{\rm n} \mathbf{I})  
= 
\varrho_{\rm n} \mathbf{g} 
+\nabla\cdot\mathbf{\Pi}_{\rm n} 
+ 
\mathbf{S_{\rm n}}
\, ,
\label{eq:neutral_momentum} 
\end{multline}
\begin{multline}
\frac{\partial E_{\rm i}}{\partial t} + 
\nabla\cdot
\left[
\left(E_{\rm i}+p_{\rm i} 
+ 
\frac{\mathbf{B}^2}{2\mu} \right)\mathbf{V}_{\rm i}-\frac{\mathbf{B}}{\mu}(\mathbf{V}_{\rm i}\cdot \mathbf{B}) 
\right] 
+ \\
\nabla\cdot
\left[
\frac{\eta}{\mu}(\nabla\times\mathbf{B})\times \mathbf{B} 
\right] 
= 
(\varrho_{\rm i} \mathbf{g} + \mathbf{S_{\rm i}}) 
\cdot \mathbf{V}_{\rm i}
+ Q_{\rm i}
+ 
\\
\nabla\cdot(\mathbf{V}_{\rm i}\cdot\mathbf{\Pi}_{\rm i}) 
+ 
\nabla\cdot{\bf q_{\rm i}} 
- L_{\rm r}^{\rm i} 
+ H_{\rm r}
\, , 
\label{eq:ion_energy} 
\end{multline}
\begin{multline}
\frac{\partial E_{\rm n}}{\partial t}+\nabla\cdot[(E_{\rm n}+p_{\rm n})\mathbf{V}_{\rm n}] 
= 
(\varrho_{\rm n} \mathbf{g} + \mathbf{S_{\rm n}}) 
\cdot \mathbf{V}_{\rm n} + Q_{\rm n} 
+ 
\\
\nabla\cdot(\mathbf{V}_{\rm n}\cdot\mathbf{\Pi}_{\rm n})
+ \nabla\cdot{\bf q_{\rm n}} 
- L_{\rm r}^{\rm n} 
\, ,
\label{eq:neutral_energy} 
\end{multline}
\begin{equation}
E_{\rm i} = \frac{\varrho_{\rm i}\mathbf{V}_{\rm i}^2}{2} 
+ \frac{p_{\rm i}}{\gamma -1 } 
+ \frac{{\mathbf B}^2}{2\mu}
\, , 
\end{equation}
\begin{equation}
E_{\rm n} = \frac{\varrho_{\rm n}\mathbf{V}_{\rm n}^2}{2} + 
\frac{p_{\rm n}}{\gamma -1 }
\, , 
\end{equation}
\begin{equation}
\frac{\partial \mathbf{B}}{\partial t}
=
\nabla \times (\mathbf{V_{\rm i} \times }\mathbf{B}- \eta \nabla\times\mathbf{B})
\, ,
\hspace{6mm}
 \nabla \cdot{\mathbf B} = 0
 \, .
\label{eq:ions_induction} 
\end{equation}
%
%
%
%
%
Here, the reaction rates of the electron impact ionization and 
radiative recombination, $\Gamma^{\rm ion,rec}_{\rm i,n}$, 
momentum collisional, $\mathbf{S_{\rm i,n}}$, 
and energy source, $Q_{\rm i,n}$, terms are 
defined as 
%
\begin{equation}\label{eq:Gamma}
 \Gamma^{\rm ion}_{\rm i} = 
-\Gamma^{\rm ion}_{\rm n}=n_{\rm n}\nu^{\rm ion}
\, , 
\hspace{5mm} 
 \Gamma^{\rm rec}_{\rm n}= 
-\Gamma^{\rm rec}_{\rm i}=n_{\rm i}\nu^{\rm rec}
\, , 
\end{equation}
\begin{equation}
\mathbf{S_{\rm i}} = 
\mathbf{R^{\rm in}_{\rm i}} 
+ 
\Gamma^{\rm ion}_{\rm i}m_{\rm i}\mathbf{V}_{\rm n} 
-
\Gamma^{\rm rec}_{\rm n}m_{\rm i}\mathbf{V}_{\rm i} 
\, ,
\end{equation}
\begin{equation}
\mathbf{S_{\rm n}} = 
-\mathbf{R^{\rm in}_{\rm i}} 
+ 
\mathbf{R^{\rm ne}_{\rm n}} 
- 
\Gamma^{\rm ion}_{\rm i}m_{\rm i}\mathbf{V}_{\rm n} 
+
\Gamma^{\rm rec}_{\rm n}m_{\rm i}\mathbf{V}_{\rm i} 
\, ,
\end{equation}
%
\begin{equation}
\mathbf{R^{\rm kl}_{\rm k}}= 
\varrho_{\rm k} 
\nu_{\rm kl}
(\mathbf{V_{\rm l}}-\mathbf{V_{\rm k}})
\, , 
\hspace{2mm} k,l=\{i,n\}\, ,
\hspace{2mm} 
l\ne k\, , 
\end{equation}
%
\begin{equation}
Q_{\rm i} = \frac{1}{2}m_{\rm i} 
\left( 
\Gamma^{\rm ion}_{\rm i}V_{\rm n}^2 
-
\Gamma^{\rm rec}_{\rm i}V_{\rm i}^2
\right)
+ 
\frac{m_i}{m_n}Q^{\rm ion}_{\rm n} 
- Q^{\rm rec}_{\rm i}
+ 
Q^{\rm in}_{\rm i}
\, ,
\end{equation}
\begin{equation}
Q_{\rm n} = \frac{1}{2} m_{\rm i}
\left( 
 \Gamma^{\rm rec}_{\rm n} V_{\rm i}^2
-
 \Gamma^{\rm ion}_{\rm n} V_{\rm n}^2 
\right)
+ Q^{\rm rec}_{\rm i} 
- Q^{\rm ion}_{\rm n} 
+
Q^{\rm ni}_{\rm n}
\, 
\end{equation}
with the chemical reactions, 
\begin{equation} 
  Q^{\rm ion}_{\rm n} = 
 \frac{3}{2} \Gamma^{\rm ion}_{\rm i} k_{\rm B}T_{\rm n}
 \, ,
 \hspace{4mm}
 Q^{\rm rec}_{\rm i} = 
 \frac{3}{2} \Gamma^{\rm rec}_{\rm n} k_{\rm B}T_{\rm i}
 \, ,
\end{equation}
and 
the collisional energy exchange terms \citep{1986MNRAS.220..133D}, 
\begin{multline} 
 Q^{\rm kl}_{\rm k} = 
 \frac{1}{2}\nu_{\rm kl}\varrho_{\rm k} 
(\mathbf{V_{\rm k}}-\mathbf{V_{\rm l}})^2 
+
\frac{3}{2}
\frac{k_{\rm B}\nu_{\rm kl}\varrho_{\rm k}}
{m_{\rm k}+m_{\rm l}}
(T_{\rm l}-T_{\rm k}) \; , 
\hspace{1mm} 
{\rm k,l}=\{\rm i,
n\}\, ,
\hspace{1mm} 
l\ne k\, . 
\end{multline}
In the above equations, 
$\mathbf{g} = [0, -g, 0]$ denotes the gravity with 
$g = 274.78$\,m\,s$^{-2}$, 
$\mathbf{B}$ is the magnetic field 
and $\mu$ is the magnetic permeability of the medium. 

The viscous stress tensor is given as \citep{Braginskii1965}
\begin{multline} 
\mathbf{\Pi}_{\rm i,n} = 
\nu_{1i,n}
\left[
\nabla {\bf V}_{\rm i,n} + 
(\nabla {\bf V}_{\rm i,n})^T
\right]
+
\left(
\nu_{2i,n}-\frac{2}{3}\nu_{1i,n} 
\right)
\nabla\cdot{\bf V}_{\rm i,n}
\end{multline}
with coefficients $\nu_{1i,n}$
and $\nu_{2i,n}$ being 
the first (shear) 
and second (bulk) parameter of viscosity, respectively. 
Here one follows \citet{Hollweg1986ApJ...306..730H} and takes 
\begin{equation} 
\nu_{1i,n} = 10^{-16}\, T_{\rm i,n}^{5/2} \hspace{2mm} 
{\rm g}\, {\rm cm}^{-1}\, {\rm s}^{-1}\, . 
\end{equation}
Additionally, for simplicity reasons $\nu_{2i,n}=0$ is set. 

The magnetic resistivity coefficient, $\eta$, 
is taken in its simplified form as \citep{Ballesteretal2018}
%
%
%
\begin{equation} 
\eta = \frac{\varrho_{\rm i}\nu_{\rm ei}+
             \varrho_{\rm n}\nu_{\rm en}
            }
{e^2 n_{\rm e}^2}\, ,
\end{equation}
where $\nu_{\rm en}$ and $\nu_{\rm ei}$ are respectively 
the electron-neutral and electron-ion collisions frequencies. 

The collision 
frequency between 
protons+electrons and 
hydrogen atoms 
is 
specified 
as 
\citep{Braginskii1965,Goodman2004A&A...416.1159G,khomenko2012ApJ...747...87K,Ballesteretal2018}
\begin{multline}
 \nu_{\rm kl} = 
  \frac{4}{3} 
 \frac{\sigma_{\rm kl}\varrho_{\rm l}}{m_{\rm k}+m_{\rm l}}
 \sqrt{ \frac{8k_{\rm B}}{\pi} 
\left(
\frac{T_{\rm k}}{m_{\rm k}}+\frac{T_{\rm l}}{m_{\rm l}}
\right) }
      \, ,
\hspace{1mm} 
{\rm k,l=\{
i,n\}}\, ,
\hspace{1mm} 
{\rm l
\ne k}
\end{multline}
with  
$\sigma_{\rm kl}=\sigma_{\rm lk}$ being 
the collisional cross-section for k- and l-species  
for which its 
classical 
values of 
$\sigma_{\rm in}=\sigma_{\rm ei}=
1.4\times 10^{-19}$~m$^{2}$ 
and 
$\sigma_{\rm en}=2\times 10^{-19}$~m$^{2}$ 
are chosen from \cite{VranjesKrstic2013}. 
See \citet{Wargnier_2022} 
for 
recently derived expressions for collision frequencies. 

The temperatures are given by the ideal gas laws, 
\begin{equation}
p_{\rm k}=\frac{k_{\rm B}}{m_{\rm k}}
\varrho_{\rm k}T_{\rm k}\, , 
\hspace{4mm} 
{\rm k}=\{\rm i,n\}\, ,
\label{eq:pressures}
\end{equation}
with the ${\rm k}$-specie gas pressure $p_{\rm k}$ and 
mass $m_{\rm k}$, 
$k_{\rm B}$ is the Boltzmann constant, 
and $\gamma=5/3$ is the specific heats ratio. 

In Eq.~(\ref{eq:Gamma}) 
the symbols $\nu^{\rm ion}$ and $\nu^{\rm rec}$ denote 
the ionization and recombination frequencies, 
i.e. 
\citep{Voronov1997,Smirnov2003,Ballai2019FrASS...6...39B,PopescuBraileanuetal2019,Snow2021MNRAS.506.1334S}:
\begin{equation}
\nu^{\rm ion}   \approx 
n_{\rm e}A\frac{1}{X+\phi_{\rm i}/T_{\rm e}^*} 
\left(\frac{\phi_{\rm i}}{T_{\rm e}}\right)^K 
\exp{\left\{-\left(
\frac{\phi_{\rm i}}{T_{\rm e}}
\right)\right\}} 
\, ,
\end{equation}
\begin{equation}
\nu^{\rm rec} \approx 
2.6\times 10^{-19}\times 
\frac{n_{\rm e}}{\sqrt{T_{\rm e}^*}}\, ,
\end{equation}
with $\phi_{\rm i}=13.6$~eV, $n_{\rm e}$ electron particle density, 
$T_{\rm e}^*$ electron temperature expressed in eV, $A=2.91\times 
10^{-14}$, $K=0.39$ and $X=0.232$. 
According to \citet{carlsson2012A&A...539A..39C} 
radiative recombination may be important in the chromosphere 
and in the low corona. 
Note that an advanced 
multi-fluid model of the solar atmosphere was recently developed 
by \citet{Martinez-Sykora2020ApJ...900..101M} which is also 
capable of treating 
nonequilibrium ionization, radiation, thermal conduction, 
and other complex processes in the solar atmosphere. 

%

The radiative loss terms, $L_{\rm r}^{\rm i,n}$, are  implemented: (a) in the photosphere and 
in the low chromosphere 
in the framework of thick radiation 
for protons+electrons and neutrals, described 
in details by 
\citet{AbbettFisher2012} 
and (b) in the higher atmospheric layers 
as thin radiation for ions \citep{MooreFung1972}. 
Note that radiation for neutrals is neglected in high atmosphere 
due to low mass density of neutrals there. 
The thick radiation for ions and neutrals are conditionally 
implemented in the solar atmosphere, for $y\ge 0$\, Mm and 
for $0.1 < \tau < 10$, 
where $\tau$ is optical distance \citep[e.g.,][]{AbbettFisher2012}. 
Otherwise, in the top chromosphere and in the solar corona, 
for $\tau\le 0.1$, thin cooling is adopted for ions. 

The symbols ${\bf q}_{\rm i,n}$ denote 
thermal conduction fluxes. For neutrals 
thermal conduction flux is isotropic and expressed 
by the following formula: 
\begin{equation}
{\bf q}_{\rm n} = \kappa_{\rm n} \nabla T_{\rm n}\, . 
\end{equation}
Here the conduction coefficient is given as 
\citep{Cranmer2007ApJS..171..520C} 
\begin{equation}
\kappa_{\rm n} = 
\frac{29.6\, T_{\rm n}}{1+\sqrt{7.6\cdot 10^5\, {\rm K} /T_{\rm n}}} 
\frac{m_{\rm n}}{k_{\rm B}}\, , 
\end{equation}
where $k_{\rm B}=1.3807\cdot 10^{-16}\, {\rm cm}^2\, 
{\rm g}\, {\rm s}^{-2}\, {\rm K}^{-1}$ is the Boltzmann constant. 
Thermal conduction for ions is strongly anisotropic 
with thermal conduction across magnetic field lines being 
negligibly small. Therefore, it is assumed that 
the thermal conduction operates along magnetic field lines 
and the flux is described as follows: 
\begin{equation}
{\bf q}_{\rm i} = \kappa_{\parallel}{\bf b} 
\nabla({\bf b}\cdot T_{\rm i}) \, , 
\end{equation}
with ${\bf b}={\bf B}/B$ being a unit vector along magnetic 
field. The parallel thermal conduction coefficient, 
$\kappa_{\parallel}$,
is 
taken from \cite{Spitzer1962} as 
\begin{multline}
\kappa_{\parallel} \approx 4.6\cdot 10^{13} 
\left(
\frac{T_{\rm e}}{10^8\, {\rm K}}\right)^{5/2}
\frac{40}{\Lambda}\, 
\hspace{3mm}
{\rm erg}\, {\rm s}^{-1}\, {\rm cm}^{-1}\, {\rm K}^{-1}  
\end{multline}
with the quantum Coulomb logarithm \citep{Honda2013JaJAP..52j8002H} 
\begin{equation}
\Lambda \approx 30.9 - \log\frac{n_{\rm e}^{1/2}}{T_{\rm e}k_{\rm B}^*}\, .
\end{equation}
Here 
$k_{\rm B}^*$ 
is 
the Boltzmann constant expressed in ${\rm eV}\, {\rm K}^{–1}$. 

In Eq.\;(\ref{eq:ion_energy}) the heating term, $H_{\rm r}$, 
is optionally set. The source of this term could be associated with 
high-frequency ion-cyclotron waves 
that operate in the upper parts of the solar atmosphere  \citep{Squire2022}, 
torsional Alfv\'en waves 
\citep{2022arXiv220702878F} 
or with any other heating process  \citep[e.g.,][]{DePontieu2022ApJ...926...52D}. 
The following cases are considered here: 
(a)~no heating with $H_{\rm r}=0$; 
(b)~heating with $H_{\rm r}=-L_{\rm r}$. 
Hence, the heating term, if adopted,  
balances 
the thin radiation and it is implied in all regions 
in which ion temperature is higher than $15\cdot 10^3\;$K. 
This value of the ion temperature corresponds to the low corona, 
and it has been chosen somehow arbitrary. 
In future studies, more realistic heating terms may be adopted 
such as, for instance, the recently used heating term 
which could be 
parameterised by a power-law function of 
the local plasma conditions, 
$H_{\rm r}\sim \varrho_{\rm i}^aT_{\rm i}^b$, 
where $a$ and $b$ are treated as free parameters 
\citep{kolotkov2022MNRAS.514L..51K}. 

To avoid the generation of transients, 
all non-adiabatic and non-ideal terms 
are ramped by setting them equal to $0$ at $t=0\;$s 
and then they are allowed to linearly grow in time 
untill $t=10^3\;$s. 
Later, they are kept equal to their physical values. 
The 
selenoidal condition of Eq.~(\ref{eq:ions_induction}) 
is controlled by a hyperbolic divergence-cleaning
technique of \citet{Dedneretal2002}. 
A second-order spatially accurate Godunov-type method 
with HLLD Riemann solver \citep{MiyoshiKusano2005} 
and 
third-order 
Runge-Kutta method 
\citep{durran2010} 
for integration in time 
with the Courant-Friedrichs-Lewy number equal to $0.9$ 
are used. 
All non-ideal and non-adiabatic terms in 
the two-fluid equations are treated implicitly in 
a separate step using operator splitting with 
Super-Time-Stepping 
technique 
\citep{https://doi.org/10.1002/(SICI)1099-0887(199601)12:1<31::AID-CNM950>3.0.CO;2-5}. 



Note that in the limit of long wavelength/period waves, 
the two-fluid equations approach the two-species equations. 
In this limit 
${\bf V}_{\rm i} \approx {\bf V}_{\rm n}$, and 
one momentum equation together with two mass conservation 
equations 
are required; e.g., one mass conservation equation for $\varrho_{\rm i}$ 
and another one for $\varrho_{\rm i}+\varrho_{\rm n}$. Such a set of equations 
is called the two-species equations 
which are widely used in space weather 
\citep[e.g.][ and references therein]{Tanaka1997JGR...10219805T, Terada2009JGRA..114.9208T,Ma2013JGRA..118..321M,Shou2016ApJ...833..160S}.
MHD equations would result from the two-species equations 
for $\varrho_{\rm n}=0$, corresponding to 
a fully-ionized medium. 
Consequently, a two-fluid model exhibits a potential implication 
in the given scientific context, 
even if it is run for long wavelength/period waves. 
Moreover, a two-fluid model is superior over an MHD model with ambipolar diffusion, as the former provides 
information 
about dynamics of neutrals, while the latter suffers from 
the lack of it. We do not discuss dynamics of neutrals 
in this paper, however. 
We focus on evolution of ions and in particular on their  temperature, vertical velocities and wave-periods of the excited 
ion waves, instead. 
These ion properties consist a set of solar observables, 
while observational techniques for neutrals require 
further developments \citep{Khomenko2016ApJ...823..132K}. 
However, see \cite{Zapi_r_2022} 
for the recent report on ion-neutral 
velocity drift observed in a solar prominence. 
\section{Computational Results}
\label{sec:results}

\subsection{Numerical model and solar atmosphere structure} 
%
%
\begin{figure}[!htbp]
\begin{center}
\includegraphics[width=0.85\textwidth]{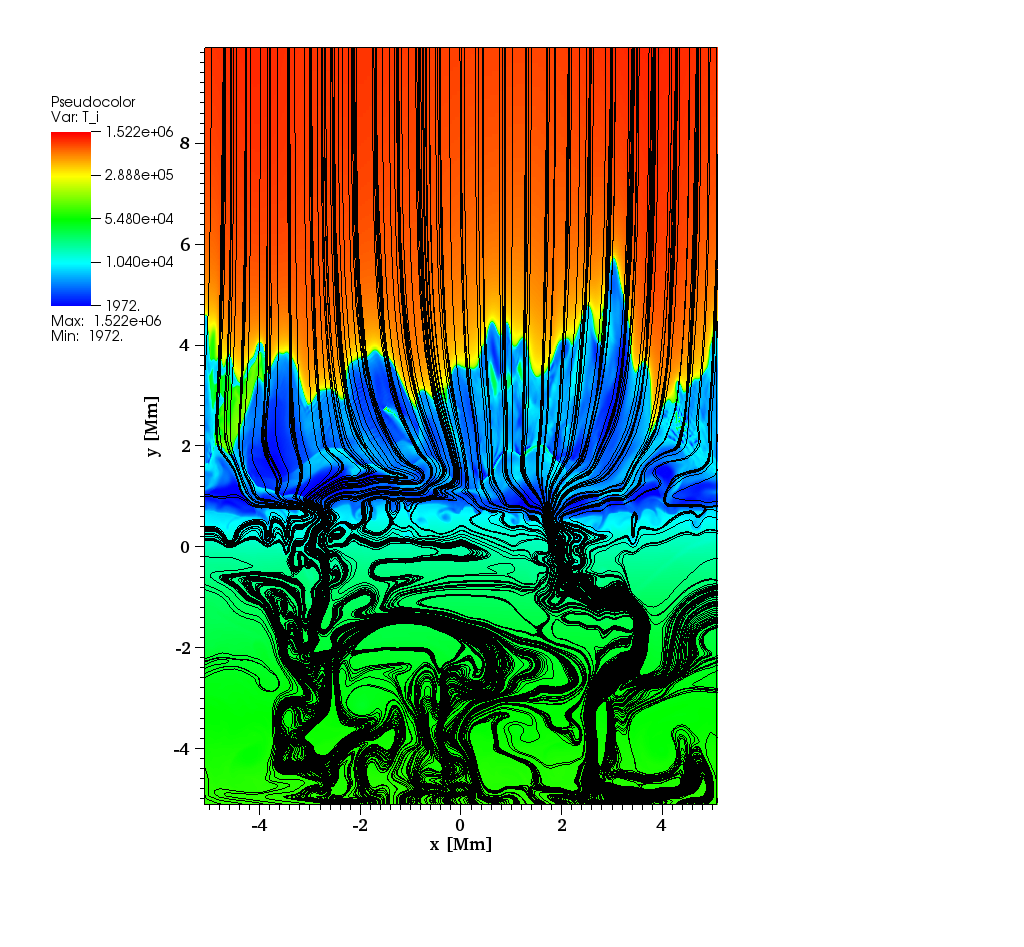}\\
\includegraphics[width=0.85\textwidth]{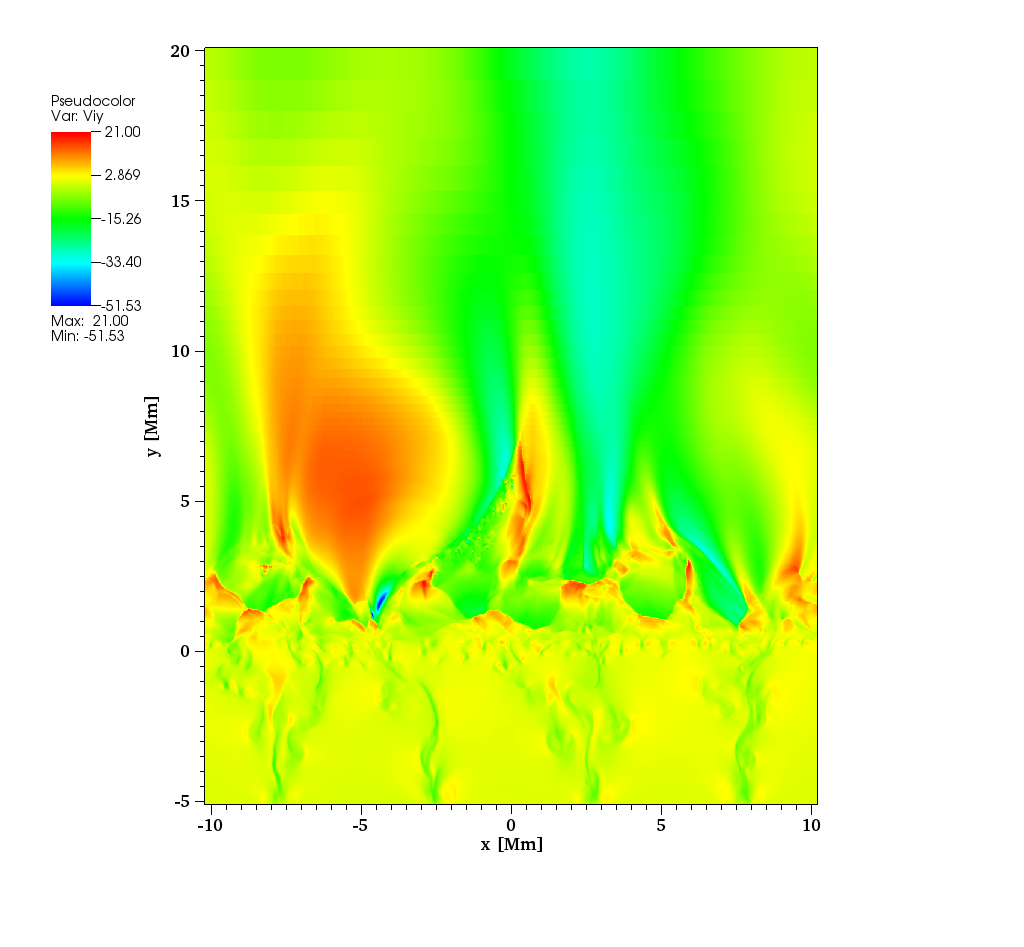}
\end{center}
\caption{Spatial profiles of $\log{T_{\rm i}}$, 
overlaid by magnetic field lines (top) 
and the vertical component of the ion velocity $V_{\rm iy}$ (bottom) 
for $H_{\rm r}=0$. 
The profiles for $H_{\rm r}=-L_{\rm r}$ 
look qualitatively similar (not shown). 
The ion temperature, $T_{\rm i}$, 
and 
vertical component of ion velocity, 
are expressed in Kelvin and km\, s$^{-1}$, respectively. 
}
\label{fig:Ti-Vi}
\end{figure}
%
%
\begin{figure*}[htbp]
\begin{center}
\hspace{-0.2cm}
\includegraphics[width=0.5\textwidth]{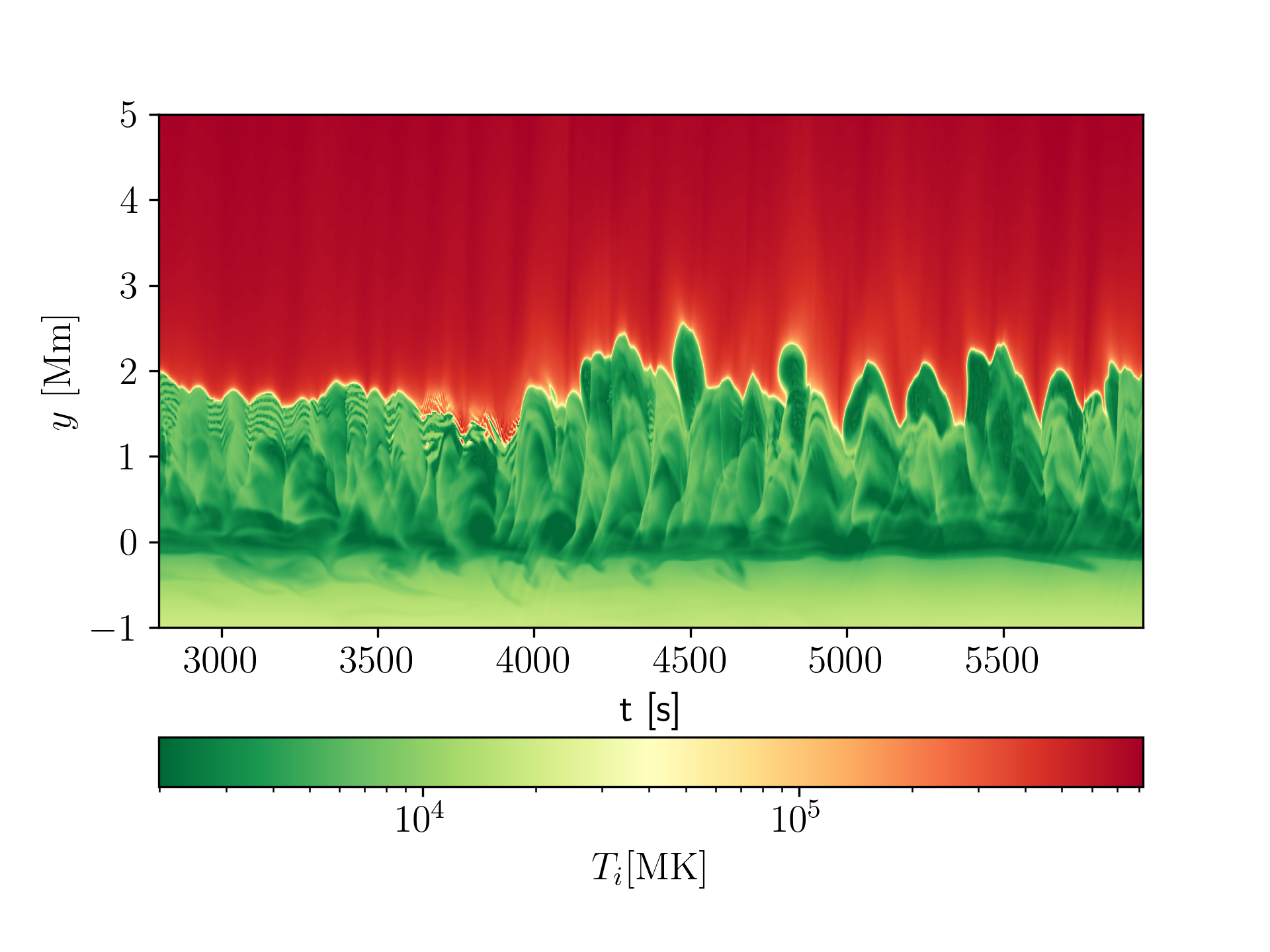}
\includegraphics[width=0.5\textwidth]{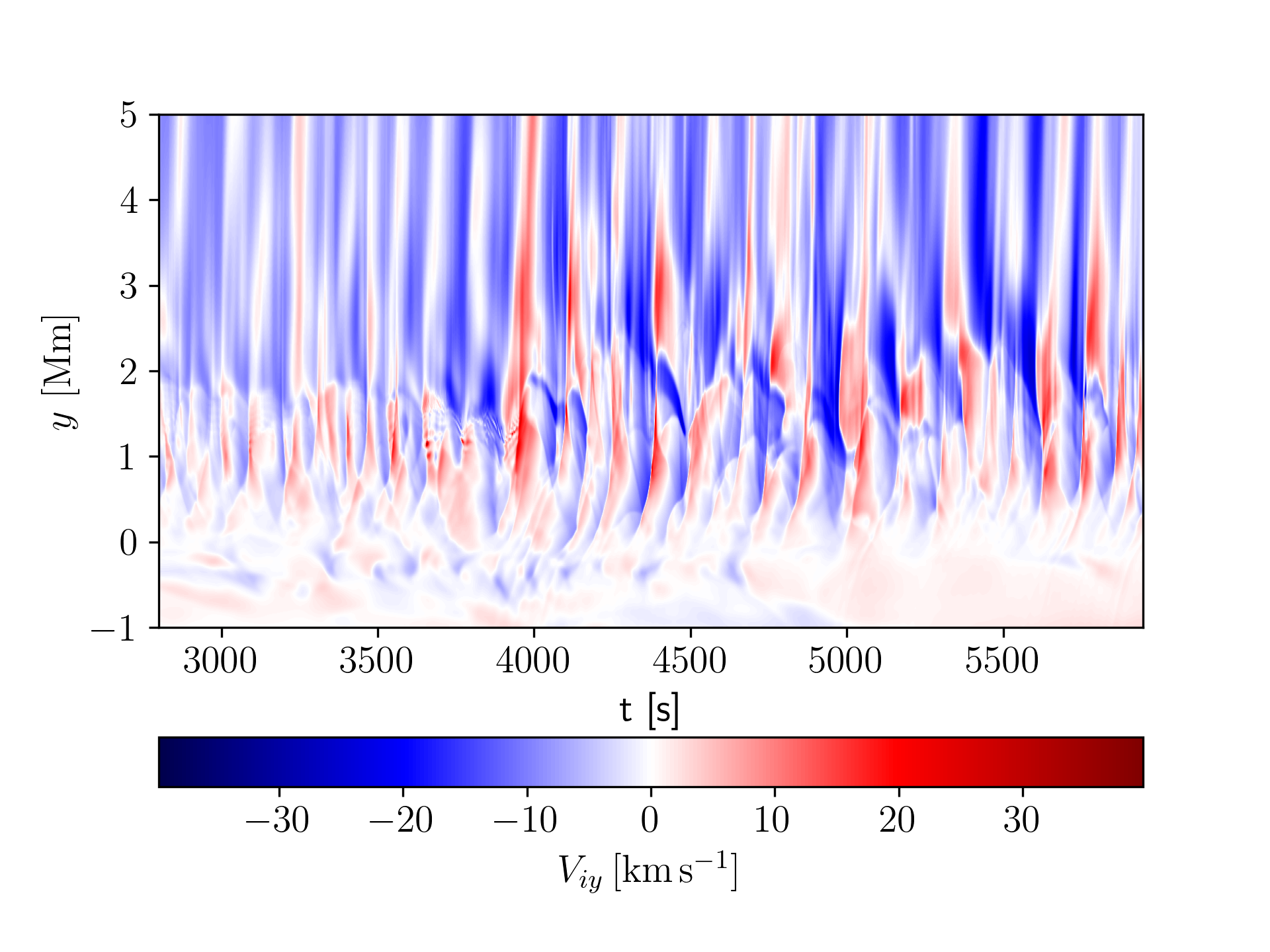}
\\
\vspace{0.25cm}
\includegraphics[width=0.475\textwidth]{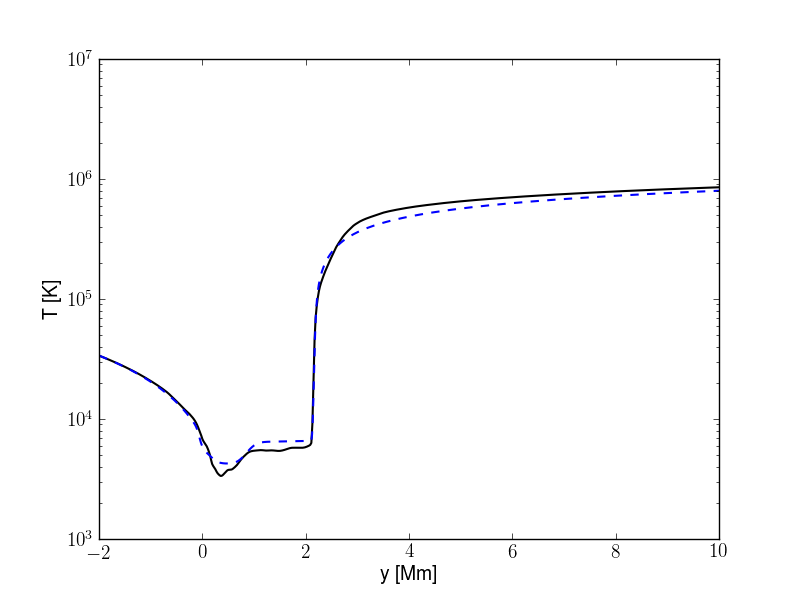}
\includegraphics[width=0.475\textwidth]{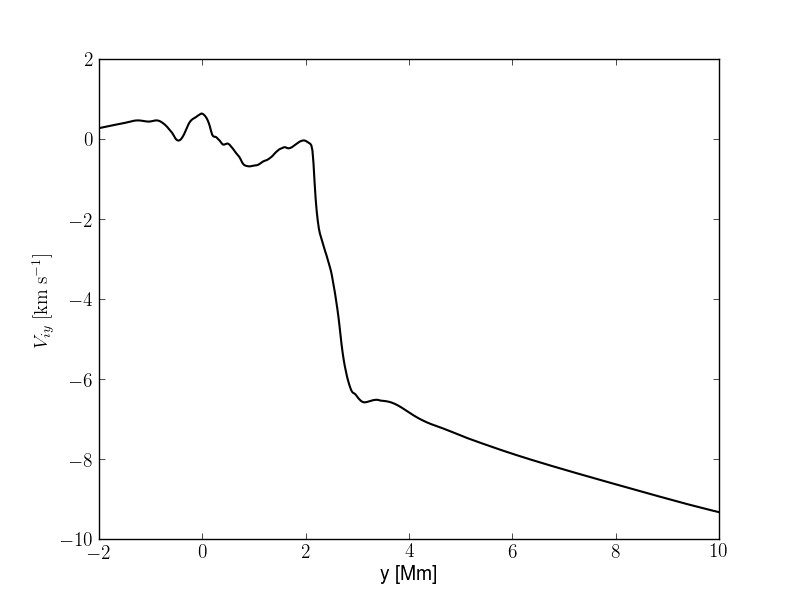}
\end{center}
\caption{Top: Time-distance plots for the ion temperature
            (left) 
            and 
            the vertical component of the ion velocity 
           (right), 
           evaluated at $x=0$. 
         Bottom: 
         Temporally averaged 
         ion temperature 
           (left: solid line), 
           semi-empirical data of \cite{AvrettLoeser2008} 
           (left: dashed line) 
            and 
            vertical component of the ion velocity (right) 
           vs height $y$ 
           for the case of $H_{\rm r}=0$.
         }
\label{fig:T-V-E}
\end{figure*}
%
%
\begin{figure*}[htbp]
\begin{center}
\includegraphics[width=0.7\textwidth]{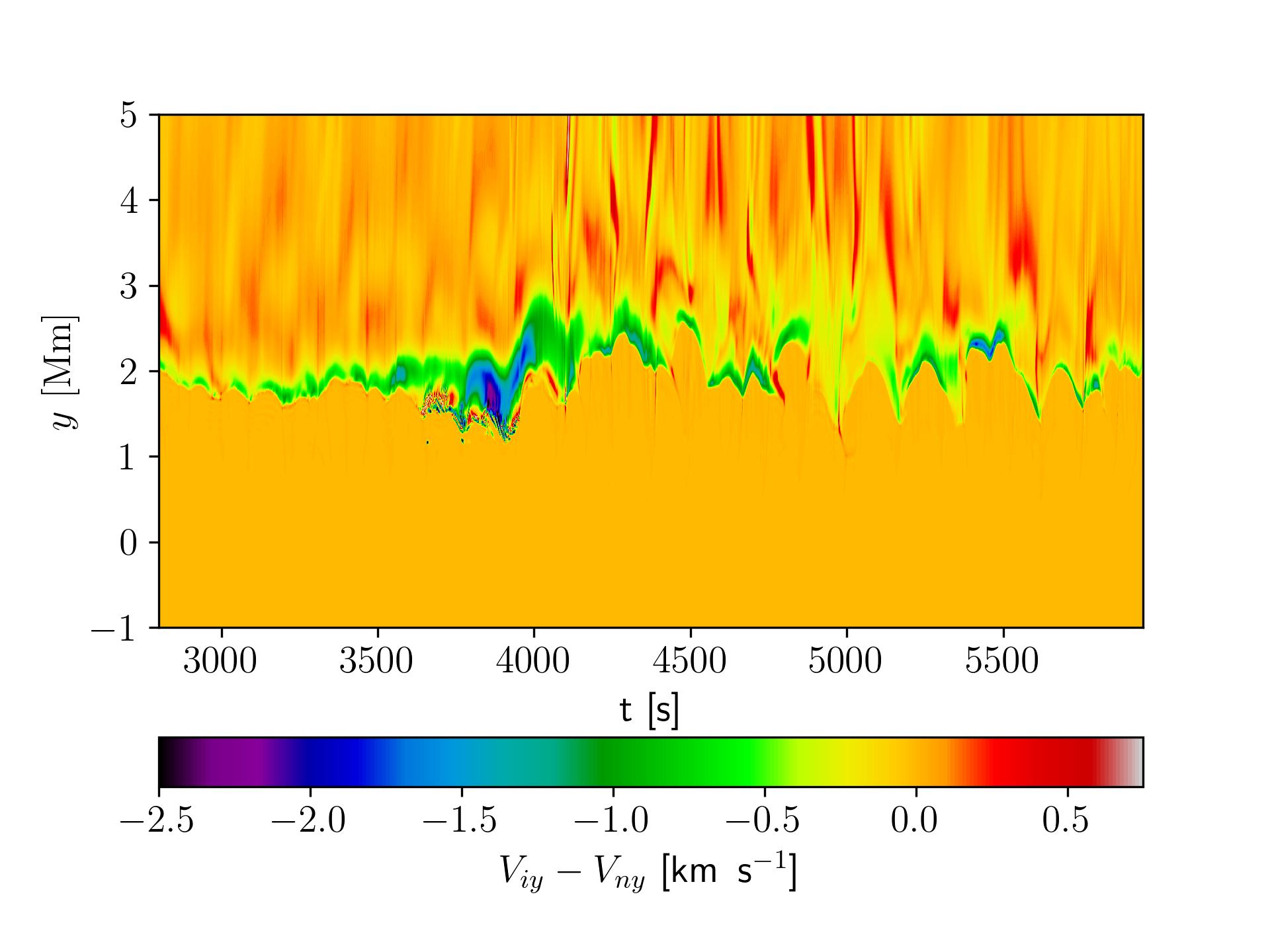}
\includegraphics[width=0.7\textwidth]{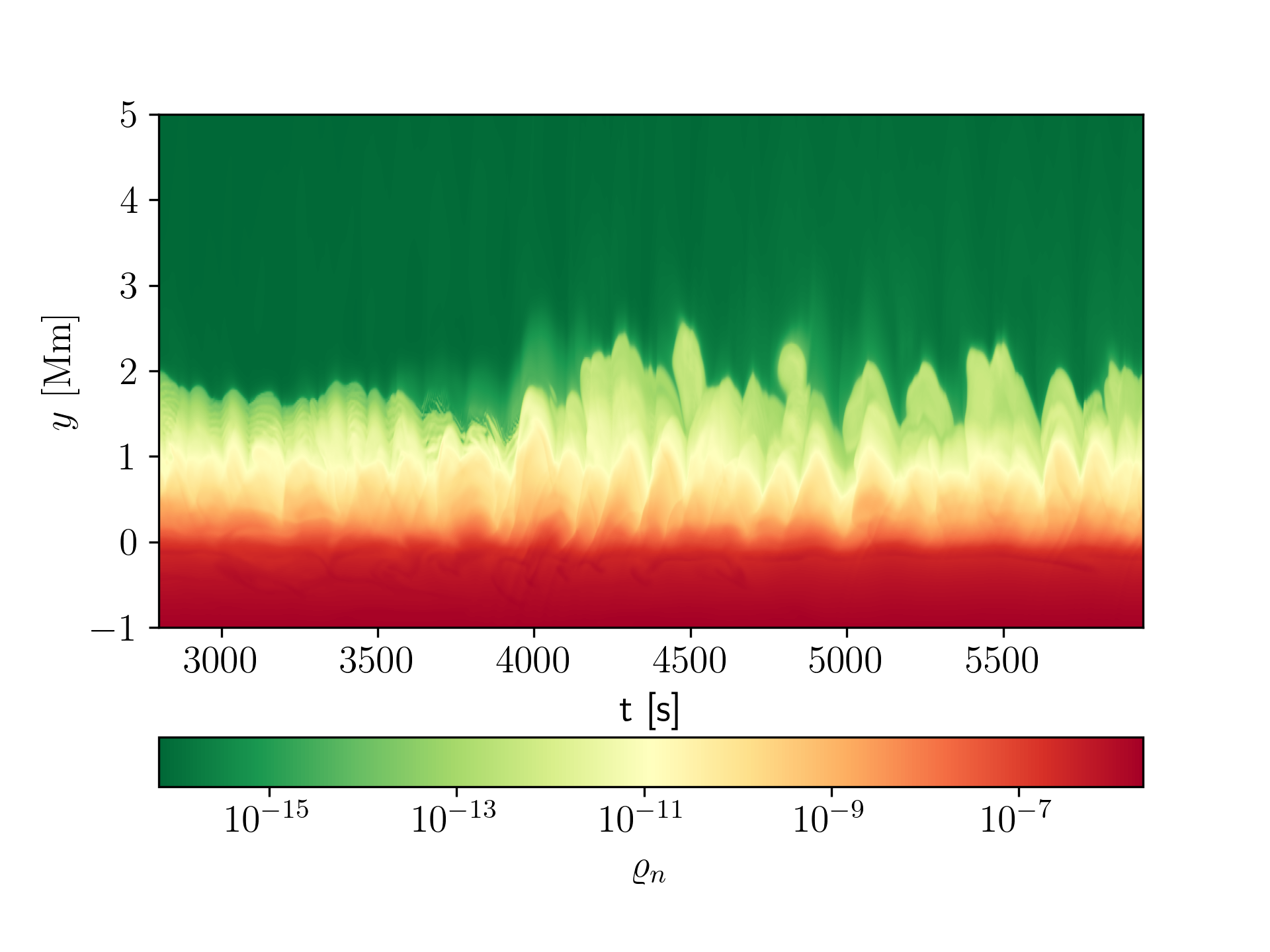}
\end{center}
\caption{Time-distance plot for the horizontally averaged 
ion and neutral vertical components of velocity drift, 
${V}_{\rm iy} - {V}_{\rm ny}$, (top) 
and neutral mass density, $\varrho_{\rm n}$, 
(bottom)
for $H_{\rm r}=-L_{\rm r}$. 
}
\label{fig:Vi-Vn}
\end{figure*}
%
%
\begin{figure}[htbp]
\begin{center}
\hspace{-0.2cm}
\\
\vspace{0.25cm}
\includegraphics[width=0.7\textwidth]{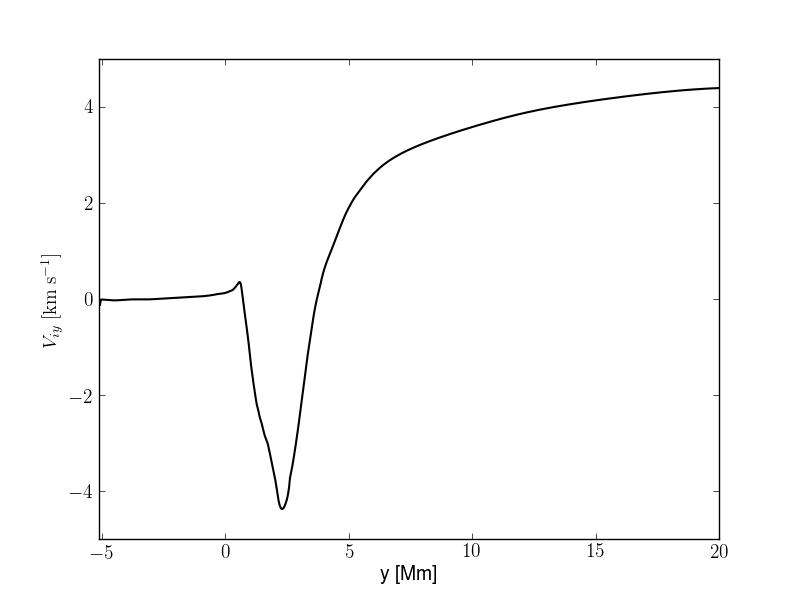}
\end{center}
\caption{Horizontally and 
         temporally averaged 
         vertical component of the ion velocity 
         vs height $y$ 
         for the case of $H_{\rm r}=-L_{\rm r}$. 
         }
\label{fig:Viy-heat}
\end{figure}
%
%
\begin{figure}[htbp]
	\centering
   \includegraphics[width=0.7\textwidth]{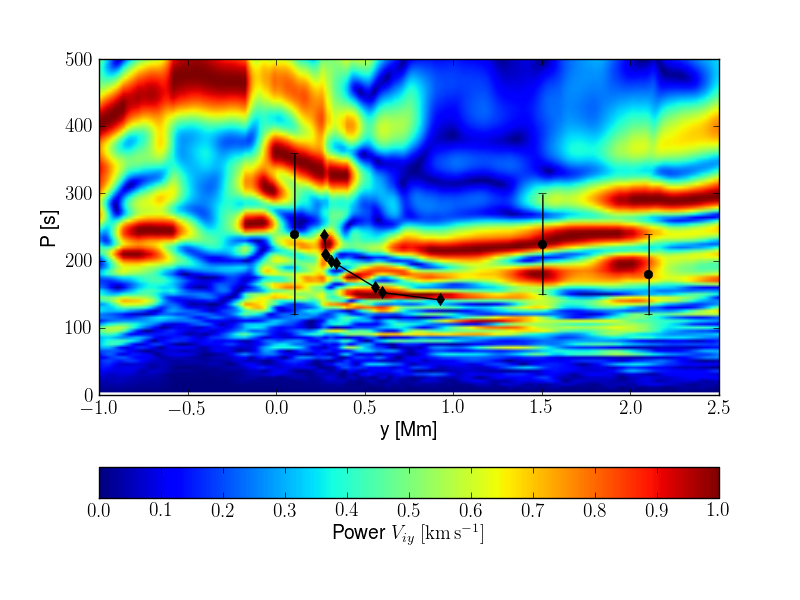}
	\caption{Wave periods, $P$, 
	evaluated from the Fourier power spectrum for the ion vertical
velocity of Fig.~\ref{fig:T-V-E} (right-top) 
(contour plots). 
The diamonds and dots show the observational data obtained by 
 \citet{Wisniewska_2016} and 
\citet{Kayshap2018}, respectively. 
}
\label{P_obs-num}
\end{figure}

The 2.5D numerical simulations of wave propagation and dissipation 
in the solar atmosphere are performed with the JOANNA code \citep{
Wojciketal2019a,Wojciketal2019b,Wojciketal2019c}, which solves 
the 
non-ideal and non-adiabatic 
two-fluid equations within 
a simulation box that is specified along the horizontal ($x$-) and vertical ($y$-) 
directions as 
$(-10.24\le x\le 10.24)$\,Mm $\times$ 
$(-5.12\le y\le 40)$\;Mm. 
The system is assumed invariant along the $z$-direction 
(i.e.\ $\partial/\partial z=0$).   
Below the level $y=5.12$\,Mm, 
a uniform grid with cell size 20\,km\, $\times$\, 20\,km 
is set, 
while higher up the grid is stretched along the $y-$direction, dividing 
it into $64$ cells whose sizes steadily grow with height. 
The stretched grid size, $\Delta y_{\rm j}$, 
is specified as
\begin{equation}
\Delta y_{\rm j} = r^{\rm j} \Delta y\, ,\hspace{3mm} 
{\rm j=1,2,\ldots, 64}\, ,
\end{equation}
where $\Delta y=20$\;km is the uniform grid size 
and the stretching ratio $r$ is given as 
\begin{equation}
y_{\rm t} - y_{\rm b} = \sum_{\rm j=1}^{j=64}\Delta y_{\rm j}\, .
\end{equation}
Here, 
$y_{\rm b}=5.12$\;Mm 
and $y_{\rm t}=40$\;Mm are the bottom-most and top-most points of 
the stretched grid zone. 

At $y=-5.12$\;Mm and at $y=40$\;Mm 
all plasma quantities are fixed to their magnetostatic values at all times $t\ge 0$\;s. 
The left and right boundary conditions are set to be periodic. 
Our simulations are initiated at $t=0$\;s by implementing a hydrostatic solar atmosphere with 
the semi-empirical temperature profile, $T(y)$, according to the model of \cite{AvrettLoeser2008}. 
This temperature, 
which initially (at $t=0$\;s) is identical for ions and neutrals, 
$T_{\rm i}(x,y,t=0)=T_{\rm n}(x,y,t=0)=T(y)$, 
uniquely determines the equilibrium 
ion and neutral 
mass 
densities 
and gas pressures \citep[e.g.][]{2020ApJ...896L...1M}. 
Then, convective instabilities 
occur in the system self-consistently. 
These instabilities are most prominent below the photosphere, 
and they lead to self-generated and self-evolving 
turbulent fields that mimic the convection with granulation 
cells at its top 
(Fig.~\ref{fig:Ti-Vi}, top). 
Such turbulent fields reshape the initial 
magnetic field, which is taken in the form of 
the four arcades, given as  
\begin{eqnarray}
    B_{\rm x} =  &B_{\rm 0}& \cos\left(\frac{x}{\Lambda_{\rm B}}\right)\,  
            \exp\left(-\frac{y}{\Lambda_{\rm B}}\right)\, , \\
    B_{\rm y} = -&B_{\rm 0}& \sin\left(\frac{x}{\Lambda_{\rm B}}\right)\,  
            \exp\left(-\frac{y}{\Lambda_{\rm B}}\right)\, , 
\end{eqnarray}
with $B_{\rm 0}=20$\;Gs, 
$\Lambda_{\rm B}=2L/\pi$ and $L=2.56$\;Mm, 
being overlaid by 
the straight magnetic field 
$[B_{\rm x},B_{\rm y},B_{\rm z}]=[0,10,2]$\;G. 
This initial magnetic field evolves 
into well developed complex structures below the transition region 
(Fig.~\ref{fig:Ti-Vi}, top). 
The spatial profile of 
$\log T_{\rm i}(x,y)$ 
exhibits a perturbed pattern 
that shows oscillations and jets in the transition region, 
which was initially located at the level of $y=2.1\;$Mm, 
as can be seen in the profile of $T(y)$ 
(Fig.~\ref{fig:T-V-E}, left-bottom, dashed line). 
The granulation-generated jets are well seen in 
the spatial profile of $\log T_{\rm i}(x,y)$; 
the largest jet is located at $x\approx 0.5$~Mm 
and it reaches the height of $y\approx 6$~Mm
(Fig.~\ref{fig:Ti-Vi}, top). 

Figure~\ref{fig:T-V-E} (left-top) shows the ion temperature, 
evaluated at $x=0$\;Mm. 
In general, its averaged-over-time quantity 
exhibits a similar distribution 
as the atmospheric temperature in the semi-empirical model of \cite{AvrettLoeser2008} (Fig.~\ref{fig:T-V-E}, left-bottom). 
\subsection{Waves and dynamics of fluid species}
\subsubsection{Chromospheric heating}
The self-generated granulation, which is responsible for the wave generation, also expels 
cold ions and neutrals from the lower atmospheric layers into the transition region 
and low corona. As a result, the transition region is shifted, which triggers plasma flows 
in the background atmosphere because the whole system is perturbed.  
The expelled ions reach their maximum 
velocities of $V_{\rm iy}\approx 21$\;km s$^{-1}$ (Fig.~\ref{fig:Ti-Vi}, bottom), 
and they are heated by ion-neutral collisions, which is an important signature in 
the context of chromospheric heating. 
As a result of ion–neutral collisions, the energy of these excited waves is dissipated. 
This dissipation is most effective for largest dispatches between ion and neutral 
velocities, and these waves may convert their energy into heat mostly in the chromosphere, 
compensating radiative and thermal losses. 
Indeed, the vertical component of the ion-neutral velocity drift, 
${V}_{\rm iy} - {V}_{\rm ny}$, 
attains largest values at the transition region and 
in the low corona (Fig.~\ref{fig:Vi-Vn}, top), 
leading to collisional heating there 
\citep{Martinez-Sykora2020ApJ...900..101M}. 

In the solar photosphere, collisions 
between neutrals and ions are frequent, and yet, wave damping is not 
significant because of the high frequency of collisions 
that equalize momenta of neutrals and ions quickly.  
On the other hand, in the solar 
chromosphere the collisions are less frequent and, as a result, 
there are differences in momenta  between neutrals and ions, 
which makes the 
damping of flows and short-wavelength waves more effective.  Our results also 
demonstrate that neutrals that reach the lower layers of the solar 
corona undergo ionization. 
The presence of such neutrals in the lower corona is 
responsible for damping of flows and 
waves that carry their 
energy up to these layers. However, the efficiency of wave damping 
in the corona is not high, therefore, the dissipated wave energy is 
not sufficient to balance the observed radiative losses \citep{Squire2022}. 
To account for these differences, an extra energy term is included in 
our numerical model (see Section~2), and the effects of this term are 
presented and discussed below. 

\subsubsection{Dynamics of neutrals}
The obtained results show that our two-fluid model reveals dynamics of 
neutrals, which play an important role in the layers of the solar 
atmosphere that are simulated in this paper. 
Figure~\ref{fig:Vi-Vn} (bottom) illustrates the time-distance 
plot of neutral mass density, $\varrho_{\rm n}$, 
collected at $x=0$\;Mm. The pattern of oscillations 
seen in this figure at the transition region, 
$y=2.1$\;Mm, is prominent and it demonstrates the role
played by neutrals in the physical processes of the solar 
atmosphere. This role can only be investigated by the two-fluid
model presented in this paper; note that no model based on MHD 
with ambipolar diffusion would give any description of the dynamics of 
neutrals. 
\subsubsection{Plasma flows}
The second central issue of the solar physics research that is addressed 
in the present paper concerns the origin of the solar wind. 
In the magnetic-free, terrestrial atmosphere the wind blows from the 
high-pressure regions to the low-pressure weather systems.  However, with the solar 
corona being permeated by magnetic fields, the nascent fast wind originates from the 
magnetic network \citep{Hassler1999}.  Moreover, \cite{Tu2005} and \cite{Tian2010} 
proposed that the wind starts in coronal funnels at altitudes in between $5$ to $20$\;Mm 
above the photosphere, and \cite{Dadashi2011A&A...534A..90D} reported 
average plasma upflows of $(-1.8 \pm 0.6)\;$km\,s$^{-1}$ 
at $1\;$MK temperature. 

From Fig.~\ref{fig:Ti-Vi} (bottom) it follows that the (red) patches of $\simeq 20$~km s$^{-1}$ 
of the ion outflows, $V_{\rm iy}$, are located at several points 
in the corona. 
The downfall of 
$\simeq -50$\;km s$^{-1}$ 
with the gravitationally attracted plasma is also clearly seen at 
a few locations. 
However, it is important that 
vertical component of ion velocity 
evaluated at $x=0$\;Mm 
exhibits quasi-periodic upflows and downfalls 
that 
are discernible at various moments in time, 
e.g.\ at $t=4\cdot 10^3$\;s 
with 
$max(V_{\rm iy}) \simeq 45$\;km s$^{-1}$ (Fig.~\ref{fig:T-V-E}, right-top), 
and the vertical component of ion velocity averaged over time 
reveals downfalls of its magnitude 
growing with height (Fig.~\ref{fig:T-V-E}, right-bottom). 
These flows 
seem to share several properties of type~I spicules 
\citep[see, e.g.,][]{Hansteen2006ApJ...647L..73H,
1986MNRAS.220..133D,
Sterling2000SoPh..196...79S,Tsiropoula2012SSRv..169..181T}. 
Note that there is some previous work on the velocity average across 
the solar atmosphere, and that the atmospheric heating occurs naturally
even within the framework of a single-fluid MHD model 
\citep[e.g.,][]{Hansteen2010ApJ...718.1070H}. 
The two-fluid model presented in this paper and the obtained 
results significantly generalize the previous work by allowing 
to describe dynamics of neutrals and ions, and their role in
the solar atmosphere heating. 

The upflows seen in the simulation are more relevant to regions of 
the quiet Sun with a vertical orientation of the magnetic field. 
Downflows are observed at the sides of the funnels with, obviously, 
oblique magnetic field. 
Regions of the quiet Sun with a horizontally 
oriented magnetic field 
do not exhibit that many upflows (Fig.~\ref{fig:Ti-Vi}, bottom).

In a quiet region, the plasma downfalls of the maximum magnitude of $5-10$\;km\,s$^{-1}$ 
and averaged upflows of about $2$\;km\,s$^{-1}$ 
were recently reported by 
\cite{Kayshap2015} and 
\cite{Tian2021arXiv210202429T}. 
The results of Figs.~\ref{fig:Ti-Vi} (bottom) and \ref{fig:T-V-E} (right), 
demonstrate that the plasma upflows 
(Fig.~\ref{fig:Viy-heat}) 
originate from the granulation-generated jets 
between $y=4$\;Mm and $y=20$\;Mm, which is consistent with the data reported  
by \citet{Tian2010}. 
Our numerical simulations show that such upflows are generated when 
the 
extra 
heating 
term in the energy equation 
is taken into account (see Section~2) to balance the radiative losses from the optically thin 
solar corona with the ion temperature $T_{\rm i} > 15\cdot 10^3$\;K 
($H_{\rm r}=-L_{\rm r}$). 

It must be pointed out that 
without the heating term ($H_{\rm r}=0$) 
only downward plasma flows 
result from our numerical simulations. 
Let us remark that the added heating term
mimics coronal plasma heating by high-frequency 
ion-cyclotron waves as recently proposed by \cite{Squire2022}. 
\subsection{Wave cutoffs and their observational verification}
Wave cutoffs arises naturally in stratified media with nonuniform magnetic fields, and they can be used to determine ranges of frequencies corresponding to propagating or evanescent waves. 
The cutoff is used to establish the ranges of periods for the propagating and reflected waves in the solar atmosphere. 
For the recent discussion 
see e.g.\ \cite{Routh2020Ap&SS.365..139R}. 
Specifically, the role of the acoustic cutoff in the solar atmosphere has been extensively studied and different formulas for this cutoff are summarized by \citet{Wisniewska_2016}, who showed that none of those formulas could reproduce their observational results. The observational results presented by \citet{Wisniewska_2016} and \citet{Kayshap2018}
demonstrated variations of the cutoff in the upper photosphere, lower chromosphere, and in the transition 
region. There have been attempts to account numerically for the observed variations of the acoustic cutoffs 
\cite[e.g,][]{Murawski2016ApJ...827...37M, Murawski2016MNRAS.463.4433M} 
but only partial agreement was found. Therefore, 
in this paper, we compute variations of the acoustic cutoff in the considered layers of the solar 
atmosphere and compare our numerical results to the observational data reported by \citet{Wisniewska_2016} 
and \citet{Kayshap2018}. 

Figure~\ref{P_obs-num} illustrates wave periods (contour plots) obtained 
from the Fourier power spectrum of $V_{\rm iy}(x=0,y,t)$, 
illustrated in Fig.~\ref{fig:T-V-E} (right-top). 
These wave periods are compared to the observational data
analyzed by \citet{Wisniewska_2016} and \citet{Kayshap2018}. 
This figure displays  
a multitude of wave power concentrations at different periods and heights, 
but a few of them correspond approximately to the location of 
the wave power concentrations found in the observational data. 
Nevertheless, the agreement between our numerical results and the
data presented in the above figure confirms that ion-neutral collisions 
are efficient energy release processes, resulting in kinetic energy 
dissipation and its conversion 
into heat. 

It must be also noted that there have been done studies of cutoffs 
of two-fluid waves in atmospheric models that have ion-neutral interactions 
included 
\citep[see references in][]{Ballesteretal2018,Alharbi2022MNRAS.511.5274A}. 
For instance, 
slow magneto-acoustic waves 
arise for sufficient short wavelengths only, 
and for long wavelengths 
these waves have only imaginary frequencies 
which correspond to non-oscillatory damping 
\citep[see Fig.~3 in][]{Zaqarashvilietal2011}. 
Similarly, according to \cite{Soler2013ApJ...767..171S} 
Alfv\'en waves of a given frequency are not propagating 
within a certain range of their wavelengths. 
%
%
\section{Conclusions and summary}
\label{sec:con}
Numerical simulations of two-fluid waves 
and plasma flows 
were performed 
in a partially ionized 
quiet-Sun region, taking into account 
non-adiabatic 
and non-ideal effects with ionization and recombination 
included self-consistently into the model \citep{Ballesteretal2018}. 
The considered 
neutral acoustic-gravity and ion Alfv\'en and magneto-acoustic-gravity 
waves were generated by spontaneously evolving and self-organizing convection. 
For the recent analysis of acoustic-gravity wave propagation 
in 3D radiation hydrodynamic numerical simulations of 
the solar atmosphere see \cite{Fleck2021RSPTA.37900170F}. 
The 
energy carried by the excited non-potential magnetic field, 
sheared plasma flows, and waves is dissipated by 
ion-neutral collisions and non-ideal (magnetic diffusivity and viscosity) effects, 
effectively heating the plasma and compensating radiative 
and thermal energy losses. 
This dissipation leads to local heating of the 
background chromosphere. 
In comparison to the previous study by 
\citet{Martinez-Sykoraetal2017}, 
\citet{Fleck2021RSPTA.37900170F}, 
\citet{Snow2021MNRAS.506.1334S}, 
and 
\citet{Navarro2022A&A...663A..96N}, 
who adopted complex non-adiabatic MHD models, 
for a partially-ionized plasma, 
and 
\citet{Wojciketal2019a} and \citet{2020ApJ...896L...1M}, who used a 
two-fluid numerical model including radiation, 
our results show that taking into account radiation, 
anisotropic thermal conduction, magnetic diffusivity, viscosity, 
ionization and recombination \citep{Ballesteretal2018} 
leads to a solar atmosphere with a vertical temperature profile 
that 
resembles the semi-empirical data of \citet{AvrettLoeser2008}.  
There were also attempts 
to assess the efficiency or feasibility of 
heating by waves by comparing the wave flux with 
the radiative loses. 
See e.g.\ \cite{Abbasvand2020ApJ...890...22A} 
for the recent studies. 
Additionally, 
the obtained 
results for wave periods show 
a quantitative agreement with the observational data of 
\citet{Wisniewska_2016} and \citet{Kayshap2018}. 

Therefore, we conclude that the granulation-generated two-fluid waves 
effectively heat the background medium 
and 
the simultaneously excited 
weak plasma outflows exhibit 
physical parameters that 
are consistent with the basic observational findings 
\citep{Hansteen2010ApJ...718.1070H,Tian2011ApJ...738...18T,Dadashi2011A&A...534A..90D}. 
To get these plasma outflows an extra heating term is required. The presence of the heating
term is evidence that the amount of energy carried by waves is not sufficient to 
heat the background atmosphere and at the same time initiate plasma outflows.  This 
limitation of the wave theory resulting from our numerical simulations is likely caused 
by the lack of momentum deposition by Alfv\'en waves, whose presence in the solar corona
is strongly confirmed by observations.  The heating term may 
actually mimic coronal heating by high-frequency ion-cyclotron waves, which was recently 
proposed by \cite{Squire2022}; however, it must be kept in mind that no plasma waves are 
considered in our numerical model. Let us also point out that the presence of these outflows 
may be responsible for the origin of the solar wind. 

To briefly summarize our work: 
the considered numerical
model and the presented results 
contribute to the studies of 
the required chromospheric 
heating and, 
in the case of heating fully balancing 
the thin cooling for $T>15\cdot 10^3$\;K, 
the origin of the fast component 
of the solar wind. 
Our present model elucidates a
general and global physical picture of the granulation-generated wave motions, plasma flows, and 
subsequent heating in the non-ideal 
quiet-Sun atmosphere. 
The improved observational estimations on such dynamical phenomena with ultra-high resolution telescopes 
(e.g., the 4m-DKIST
and the upcoming 4m-EST) may further put forward more refinement on such studies in the forthcoming time and, hence, reveal mass and energy transport processes.

%
%
\bmhead{Acknowledgments}
K.M. expresses his thanks to 
Fan Zhang, 
Teimury Zaqarashvili, 
Elena Khomenko, 
B{\l}a\.zej Ku\'zma, 
Michaela Brchnelova, 
Ramon Oliver, 
Gabor Toth, 
Takashi Tanaka, 
Naoki Terada, 
and 
Ryoya Sakata 
for 
stimulating discussions on the adopted two-fluid model. 
The JOANNA code was developed by Darek W\'{o}jcik with some 
contribution of 
Luis Kadowaki 
and 
Piotr Wo{\l}oszkiewicz.  
This work was 
done within the framework of the project from the Polish Science Center 
(NCN) Grant No. 
2020/37/B/ST9/00184. 
A. K. Srivastava acknowledges the ISRO Project Grant (DS\_2B512 13012(2)/26/2022-Sec.2) for the support of his research.
We visualize the simulation data using the VisIt software package \citep{Childs_et_al_2012}. 
SP acknowledges support from the projects
C14/19/089  (C1 project Internal Funds KU Leuven), G.0D07.19N  (FWO-Vlaanderen), SIDC Data Exploitation (ESA Prodex-12), and Belspo project B2/191/P1/SWiM.

\bibliographystyle{aa}
\bibliography{heat+wind-bib}

\end{document}